\documentclass[smallextended]{svjour3}

\smartqed

\usepackage{graphicx}
\usepackage{amsmath}
\usepackage{natbib}
\usepackage{float}

\journalname{Space Science Reviews}
\label{fig:orbit}
\label{fig:orbitmaneuvers}
\label{fig:fov}
\label{fig:gci}
\label{fig:spacecraft}
\label{fig:activity}
\begin{document}

\title{The Carruthers Mission Concept and Performance}

\author{
W.~Craig \and
T.~Immel \and
L.~Waldrop \and
J.~Troeltzsch \and
E.~Taylor \and
J.~McPhate \and
K.~Rider \and
A.~Butterworth \and
D.~Cosgrove \and
J.~Thorsness \and
H.~Filippini \and
K.~Lee \and
S.~Shomo \and
C.~McGinn \and
C.~Peterson
}

\institute{
W. Craig, T. Immel, E. Taylor, J. McPhate, K. Rider, A. Butterworth, D. Cosgrove, J. Thorsness \at University of California, Berkeley
\and
L. Waldrop, H. Filippini  \at University of Illinois
\and
J. Troeltzsch, K. Lee, S. Shomo, C. McGinn, C. Peterson \at BAE Systems, Boulder, CO
}

\date{Received: date / Accepted: date}

\maketitle

\begin{abstract}

The Carruthers Geocoronal Observatory (Carruthers), formerly GLIDE, is a NASA Heliophysics Science Mission of Opportunity implemented through the Solar Terrestrial Probes (STP) Program and launched as a rideshare in September 2025. Carruthers is the first spaceflight mission dedicated to continuous global imaging of Earth’s hydrogen exosphere through observations of geocoronal Lyman-$\alpha$ emission at 121.6 nm. The observatory operates at distances of 1.3–1.7 million km in a halo orbit about the Sun–Earth L1 point, enabling retrieval of the three-dimensional distribution of atomic hydrogen, the dominant constituent of the exosphere, on hourly timescales and unprecedented spatial resolution . These new data provide the key to understanding processes governing atmospheric escape, geospace coupling, and solar-wind interaction. Carruthers carries the GeoCoronal Imager (GCI), a dual-channel ultraviolet imaging instrument comprising the Narrow-Field Imager (NFI) for high-resolution observations of the inner exosphere and the Wide-Field Imager (WFI) for synoptic imaging of the extended hydrogen halo. Along with a student-provided experiment, the instrument suite is the sole payload aboard a three-axis-stabilized spacecraft designed to support nadir viewing throughout the $\sim178$ day halo orbit about L1. Observations also measure the interplanetary Lyman-$\alpha$ background, enabling separation of heliospheric and geocoronal emissions. The student-led experiment monitors solar Lyman-$\alpha$ and extreme ultraviolet emission during portions of the orbit. Science data are returned through the Deep Space Network at data rates up to 1 Mbit s$^{-1}$. The two-year baseline mission begins in March 2026, with propellant reserves capable of supporting more than ten years of orbit maintenance and maneuvers.

\end{abstract}

\section{Introduction}

The NASA Carruthers Geocoronal Observatory (Carruthers), formerly GLIDE, addresses key gaps in our understanding of Earth’s exosphere by establishing the first dedicated platform for continuous monitoring of exospheric hydrogen \citep{waldrop2026}. Operating in a halo orbit about the Sun--Earth L1 (L1) point at distances of 1.3--1.8 million km, Carruthers provides a stable vantage from which the global distribution of Lyman-$\alpha$ emission can be imaged without the geometric limitations inherent to near-Earth platforms. Carruthers is named after the pioneering researcher George R. Carruthers whose instruments were carried to the moon on Apollo 16 that rendered the first 2-D and spectrographic UV images of Earth \cite{carruthers76}. These revealed the extended corona of solar HI emissions at 121.6 nm that are resonantly scattered by exospheric neutral hydrogen. Measurement of this radiance distribution provides key information for the retrieval of exospheric neutral hydrogen densities. \citep{zoennchen15,rairden86}.

\begin{figure}[H]
\centering
\includegraphics[width=\linewidth]{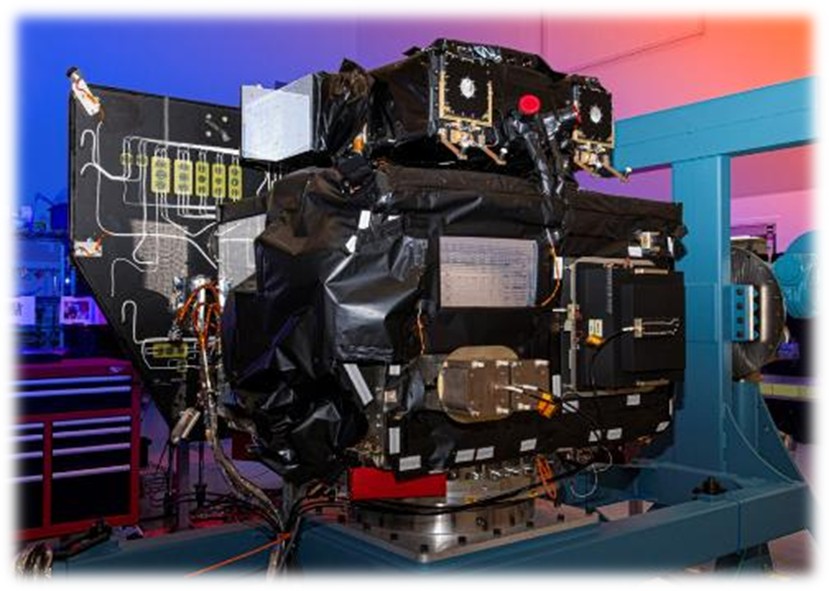}
\caption{Photograph of the Carruthers observatory showing the GCI payload atop the spacecraft bus.}
\label{fig:observatory}
\end{figure}

Observations of the geocoronal emissions of the exosphere from the lunar surface have the property that they are made completely external to the exosphere itself. This enables the most accurate retrieval of the hydrogen density. This property is also afforded to observations from L1, where the Carruthers observatory will view the geocorona at angles from the sun-Earth line between $8^\circ$ and $28^\circ$, enabling persistent measurement of the full exospheric envelope and its temporal evolution. The primary instrument is the GeoCoronal Imager (GCI), a dual-channel ultraviolet imaging system optimized for continuous high-resolution observations of the geocorona \citep{sirk2026}. The instrument is mounted atop a compact spacecraft bus (Figure~\ref{fig:observatory}) that provides the propulsion and attitude control functions required to observe the Earth, the geocorona, and stellar calibration targets.

The combined observatory delivers continuous global data sets capable of distinguishing geocoronal emission from the interplanetary Lyman-$\alpha$ background, an essential prerequisite for accurate retrievals of hydrogen density and its three-dimensional distribution. As the first mission capable of continuous global monitoring of Earth's hydrogen exosphere, Carruthers will provide new insight into the processes that couple the upper atmosphere to interplanetary space and influence the long-term evolution of Earth's atmosphere.

\section{Measurement Approach}

The Carruthers measurement approach is designed to retrieve the global, three-dimensional distribution of neutral hydrogen in Earth’s exosphere by imaging resonantly scattered solar Lyman-$\alpha$ emission at 121.6-nm from a stable vantage point near L1. This observing geometry enables nearly continuous, high-cadence global imaging of the geocorona, overcoming the spatial, temporal, and line-of-sight limitations that have constrained previous measurements from low Earth orbit or highly elliptical Earth orbits \citep{joshi2026}.

\subsection{Launch, Cruise and L1 Orbit}

\begin{figure}
\centering
\includegraphics[width=\linewidth]{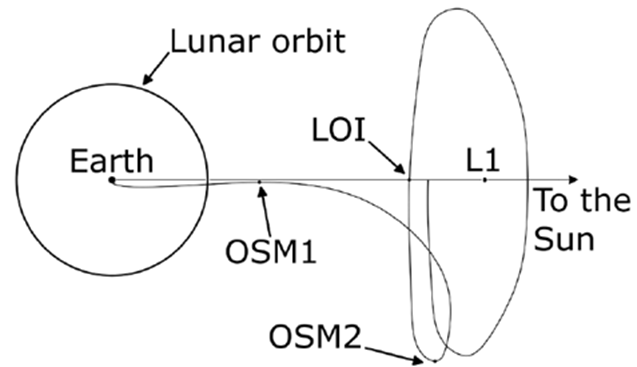}
\caption{Carruthers orbit maneuvers used to reach the operational halo orbit about Earth-Sun L1. The key maneuvers, orbit shaping maneuver (OSM) 1 and 2 and the Lissajous orbit insertion (LOI) maneuver are marked.}
\label{fig:orbitmaneuvers}
\end{figure}

Carruthers' observatory is launched as a secondary "rideshare" where the two other observatories (primary: NASA IMAP (\cite{hegarty26,mccomas18}, second secondary: NOAA SWFO-L1) are also targeting an L1 halo orbit for operations. It attains orbit about L1 by performing a series of propulsive maneuvers after separation from the launch vehicle (Figure~\ref{fig:orbitmaneuvers}). Approximately one week after separation, an initial orbit shaping maneuver (OSM\#1) corrects for small errors in the initial trajectory provided by the Falcon~9 launch vehicle. After this first maneuver, cruise phase begins, the payload ICP is turned on and the payload is warmed to "bakeout" any residual contaminant. Payload doors are opened 21 days after separation to begin commissioning of the instruments. At approximately 60 days from separation, a second maneuver places the observatory on track to intercept a point in the halo orbit about L1. This point is reached about 100 days after separation at which time the Lissajous orbit insertion takes place that inserts the spacecraft into a stable $\sim178$ day orbit about the L1 point. Stationkeeping maneuvers are planned on a 2-3 month basis after that and continue throughout the operational lifetime of the mission.

Carruthers can maintain Earth-nadir pointing throughout the science phase, viewing the exosphere at angles between approximately $8^\circ$ and $28^\circ$ from the Earth--Sun line.

\begin{figure}[H]
\centering
\includegraphics[width=\linewidth]{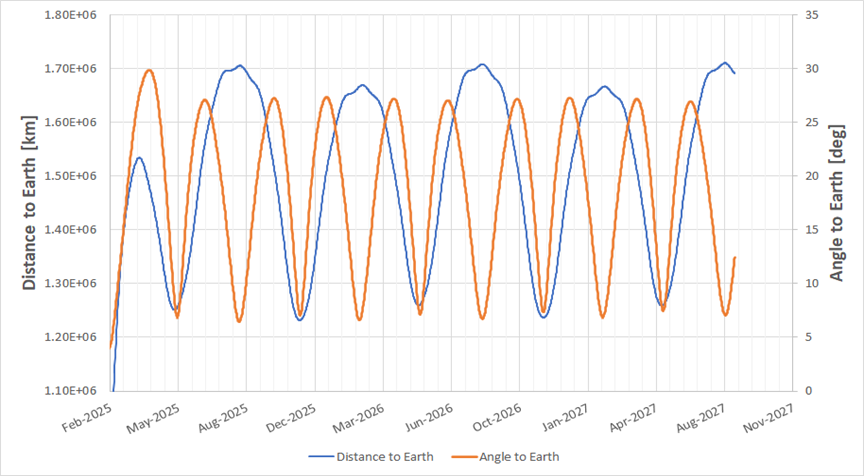}
\caption{Carruthers orbit about L1 with distance and angle with respect to Earth.}
\label{fig:halogeometry}
\end{figure}

\subsection{Dual-Channel Imaging Strategy}

The primary payload, the GeoCoronal Imager (GCI), implements a dual-channel imaging strategy using co-aligned telescopes with nested fields of view optimized for different regions of the exosphere. One channel emphasizes high spatial resolution observations of the inner exosphere, where density gradients are steep and where thermal evaporation, charge exchange, and ion--neutral coupling strongly influence structure. The second channel provides wide-angle, high-sensitivity imaging of the extended hydrogen halo, enabling characterization of large-scale morphology shaped by solar radiation pressure and interactions with the heliosphere. These channels are termed the Narrow and Wide Field Imagers (NFI and WFI). Though the reflective optical designs differ, the imagers share key design features, with identical filter wheels, ultraviolet cameras and single-shot door mechanisms.

Simultaneous operation of the two channels allows Carruthers to capture the full dynamic range of Lyman-$\alpha$ brightness in a single observation sequence. This nested approach ensures continuity between inner and outer regions of the exosphere and supports the retrieval of global hydrogen density profiles without relying on extrapolation between disparate datasets. The spatial sampling resolution and temporal cadence of the NFI and WFI are tuned to meet the science requirements of the inner and outer exosphere.

\begin{figure}
\centering
\includegraphics[width=\linewidth]{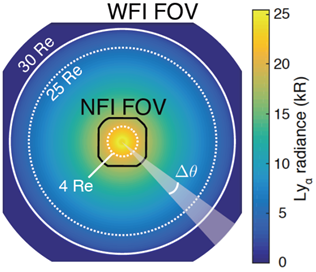}
\caption{Nested fields of view of the Near-Field Imager (NFI) and Wide-Field Imager (WFI) shown in units of Earth radius.}
\label{fig:fov}
\end{figure}

\subsection{Radiometric Measurement Concept}

Carruthers measures the Lyman-$\alpha$ radiance distribution with a target absolute accuracy of approximately 30\% within defined spatial and temporal resolution elements. The isolation of the geocoronal signal from instrumental and astrophysical backgrounds is achieved through a combination of spectral discrimination, observing geometry, and on-orbit calibration.

\subsection{Interplanetary Background Estimation}

A key advantage of the Carruthers measurement approach is the ability to measure the Ly-$\alpha$ interplanetary hydrogen (IPH) background, collect these observations into a continuous belt of observations near the ecliptic, and subtract it directly from science images. During the science phase, the interplanetary contribution is measured using the outer annulus of each wide-field image, spanning approximately $30$--$35\,R_E$.  During the initial commissioning phase, additional perimeter scans are performed to map the detectable exospheric boundary and verify the validity of using the outer WFI pixels to characterize the IPH. These observations provide additional capability toward measuring the IPH, which can be implemented as necessary to enable a valid IPH measurement free of exospheric emissions, and so the most useful correction of science images during routine operations \cite{Zhang26b}.

\subsection{Calibration and Validation Strategy}

Absolute radiometric calibration is achieved through a combination of stellar observations, filter-based spectral characterization, and detector background measurements. Bright Lyman-$\alpha$ stars allow for direct measurement of instrument responsivity at the line center, while observations of stars with strong extreme-ultraviolet continua enable characterization of the wavelength-dependent response across the instrument passband. Detailed calibration methodologies and performance are described in the companion papers in this issue \citep{Zhang26c}.

\subsection{Signal-to-Noise Performance and Retrieval Capability}

End-to-end performance analyses based on measured instrument characteristics indicate substantial margin relative to signal-to-noise and retrieval accuracy requirements across anticipated observing conditions. The measurement concepts described above are implemented in flight by the payload described in Section~\ref{sec:instrumentation} and by \cite{sirk2026}, while detailed retrieval methodologies are presented in two companion papers \citep{Joshi26a,Joshi26b}.

\section{Instrumentation}
\label{sec:instrumentation}

\subsection{Payload Architecture Overview}

The GCI payload consists of two co-aligned telescopes, ultraviolet cameras, filter wheel mechanisms, deployable doors, high-voltage power supplies, and instrument control electronics. The payload housing is an aluminum enclosure, where all of these elements are either inside the box or mounted to the outside. The enclosure is attached to the spacecraft by three bipods.  Carruthers carries a student-built solar monitor (COSSMo) that measures solar Lyman-$\alpha$ and soft X-ray irradiance to provide context for exospheric variability \citep{naldoza2026}. COSSMo is regularly exposed to sunlight, while the main instrument enclosure is typically shaded from illumination by the fixed solar panel except for flat field observations.

\subsection{Payload Mechanical Detail}

The physical realization of the Carruthers measurement strategy is the GeoCoronal Imager (GCI), a dual-channel ultraviolet imaging instrument designed to provide continuous global observations of Earth’s hydrogen exosphere in resonantly scattered solar Lyman-$\alpha$ emission. GCI consists of two co-aligned imaging telescopes—the Near-Field Imager (NFI) and the Wide-Field Imager (WFI)—mounted on a common optical bench and supported by mechanisms, electronics, and calibration subsystems. The payload design draws extensively on heritage from the ICON FUV instrument \cite{mende2017} while incorporating mission-specific adaptations required for long-duration nadir pointing from the Sun–Earth L1 environment.

\begin{figure}
\centering
\includegraphics[width=\linewidth]{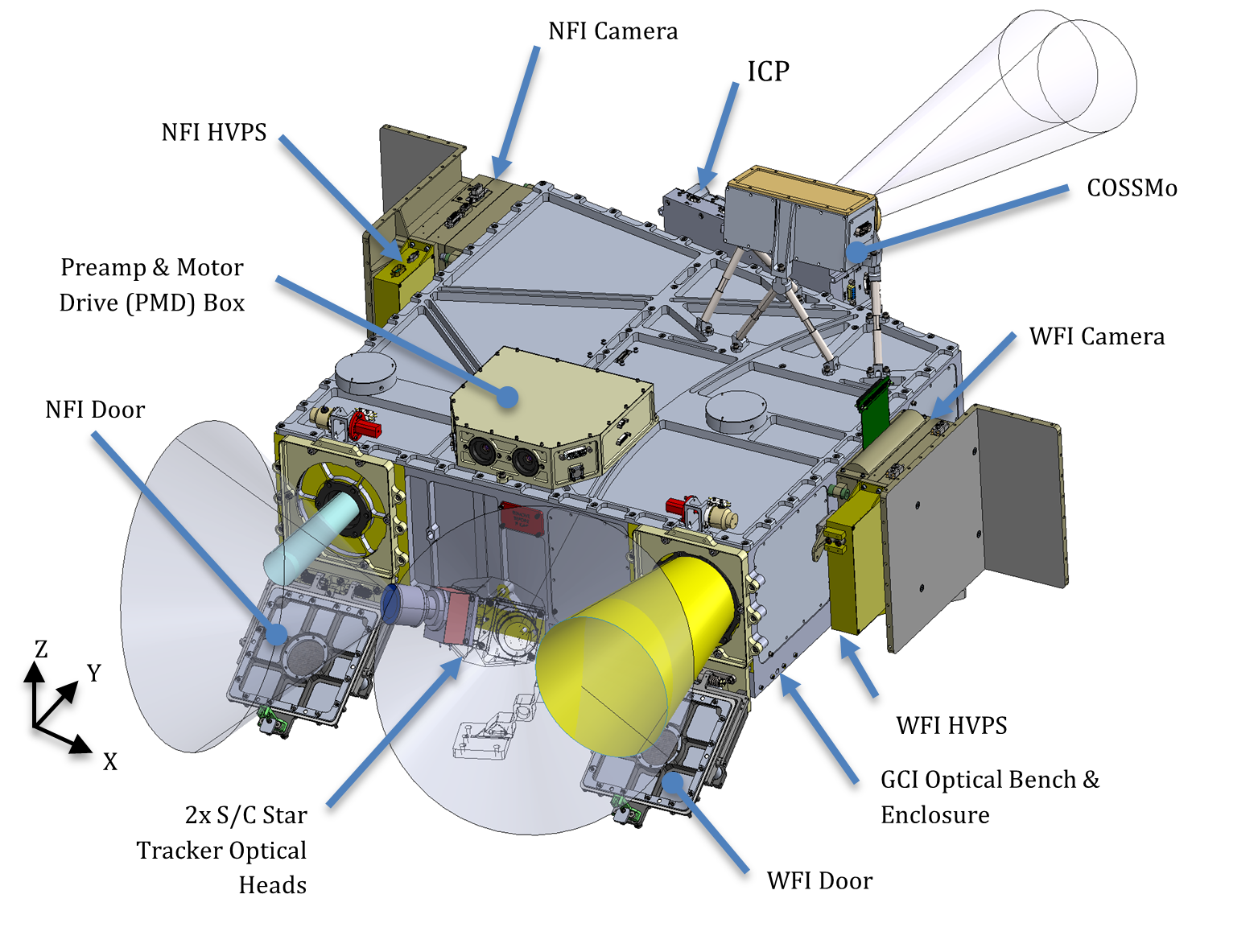}
\caption{GeoCoronal Imager payload architecture and major subsystems.}
\label{fig:gci}
\end{figure}

The GCI (Figure \ref{fig:gci})is organized around an aluminum optical bench that provides a common reference for the NFI and WFI optical trains, filter wheels, cameras, and fiducial alignment elements. The spacecraft star trackers are also mounted to the payload to maintain tight coupling between the payload boresight and the spacecraft attitude control determination.  Major payload elements include NFI and WFI telescopes, ultraviolet camera assemblies, filter wheel mechanisms, deployable doors, high-voltage power supplies, the Instrument Control Package (ICP), the Preamp and Motor Drive electronics, and thermal and contamination control hardware.

In addition to GCI, Carruthers carries a student-built solar monitor (COSSMo) that measures solar Lyman-$\alpha$ and soft X-ray irradiance to provide context for exospheric variability \citep{naldoza2026}. Unlike the GCI, its FOV is pointed away from Earth, at a median angle that places its boresight on the sun during nominal nadir viewing, where the -Z axis of the spacecraft, opposite the GCI mounting location, is exposed to sunlight. That is, the COSSMo boresight is oriented in the spacecraft Y-Z plane with a negative z component.

\subsection{Optical Design and Optomechanics}

Both NFI and WFI employ all-reflective ultraviolet optical designs optimized for Lyman-$\alpha$ imaging. The two telescopes differ in field of view and focal length to address distinct regions of the exosphere. The optical layouts are implemented using high-heritage mirror substrates with AlMgF$_2$ coatings applied at both NASA Goddard Space Flight Center and commercial vendors.

The optical bench and mirror mounts employ spherical washer stacks and precision fiducial reference surfaces to enable deterministic alignment. Alignment proceeds from individual optic characterization through full instrument integration, with wavefront error and point spread function verified at multiple stages. The final ultraviolet alignment and calibration were performed under vacuum conditions to ensure flight-like optical performance \citep{sirk2026}.

\subsection{Cameras and Ultraviolet Detectors}

Each GCI channel is equipped with an ultraviolet camera consisting of a sealed-tube image intensifier coupled through a fiber-optic taper to a solid-state focal plane array. The ultraviolet converter employs a KBr photocathode deposited on a microchannel-plate stack followed by a phosphor screen and optical taper to the detector \citep{siegmund2007}.

By optimizing the intensifier configuration and optical coupling to the CMOS detector, the system achieves angular resolution better than one arcminute across the field of view, well matched to the intrinsic spatial resolution of the microchannel plates and the detector pixel scale.

\subsection{Mechanisms: Filter Wheels and Doors}

Both NFI and WFI employ identical six-position filter wheels driven by stepper motors with heritage from the AWE \citep{lamborn2024} and ICON FUV \citep{mende2017} missions. Filter positions include nominal Lyman-$\alpha$ science filters, out-of-band filters used for background characterization, and open and blocked positions.

Deployable instrument doors protect the optics during launch and early mission phases. The door mechanisms provide one-time actuation using pin-pullers, spring-driven deployment, and end-of-travel damping to control shock during opening.

\subsection{Instrument Electronics}

The GCI electronics architecture consists of three primary subsystems: the Instrument Control Package (ICP), the Preamp and Motor Drive electronics, and the High Voltage Power Supplies (HVPS). The ICP serves as the instrument computer and power distribution hub, providing command and telemetry interfaces with the spacecraft, and managing science data handling. The PMD electronics provide stepper motor drive for the filter wheels and preamplification for sun-sensor photodiodes. The HVPS units supply high voltage to the ultraviolet intensifiers and are derived from heritage designs developed for the ICON mission \citep{immel2018}. The ICP controls the filter wheels and the cameras of the GCI and co-adds pixels to match a selected resolution. It polls the spacecraft for timestamps and pointing quaternions determined by the spacecraft attitude determination and control system. All of the camera, filter wheel and HVPS states are included in the telemetry stream, along with the pointing quaternion at the start and end of each camera integration as well as the spacecraft position held by the onboard ephemeris. This last innovation simplifies the science data pipeline, and eliminates the need for a separate ancillary data product to interpret the images.

\subsection{Thermal Control and Contamination Management}

Thermal control of the payload is achieved through a combination of passive design features and active heater control. External surfaces are protected with multilayer insulation, while the internal optical surfaces are black anodized to suppress stray light and control radiative coupling.

Contamination control measures included molecular adsorber coatings, purge hardware, vacuum bake-outs of sensitive components, and controlled integration and test procedures. These measures preserve ultraviolet throughput throughout the mission lifetime.

\subsection{Payload Integration, Alignment, and Test}

The integration and test of the payload followed a deterministic approach beginning with the characterization at the optical level and progressing through full instrument assembly, alignment, and environmental testing. Initial optical alignment was performed at the UC Berkeley Space Sciences Laboratory, while final ultraviolet calibration was conducted at the Centre Spatial de Liège using vacuum collimator facilities.

Post-calibration vibration testing was followed by re-verification of instrument performance in vacuum conditions. No adjustment of the instrument focus was required following the final alignment of the payload.

\section{Spacecraft}

\subsection{Overview and Design Heritage}

\begin{figure}
\centering
\includegraphics[width=\linewidth]{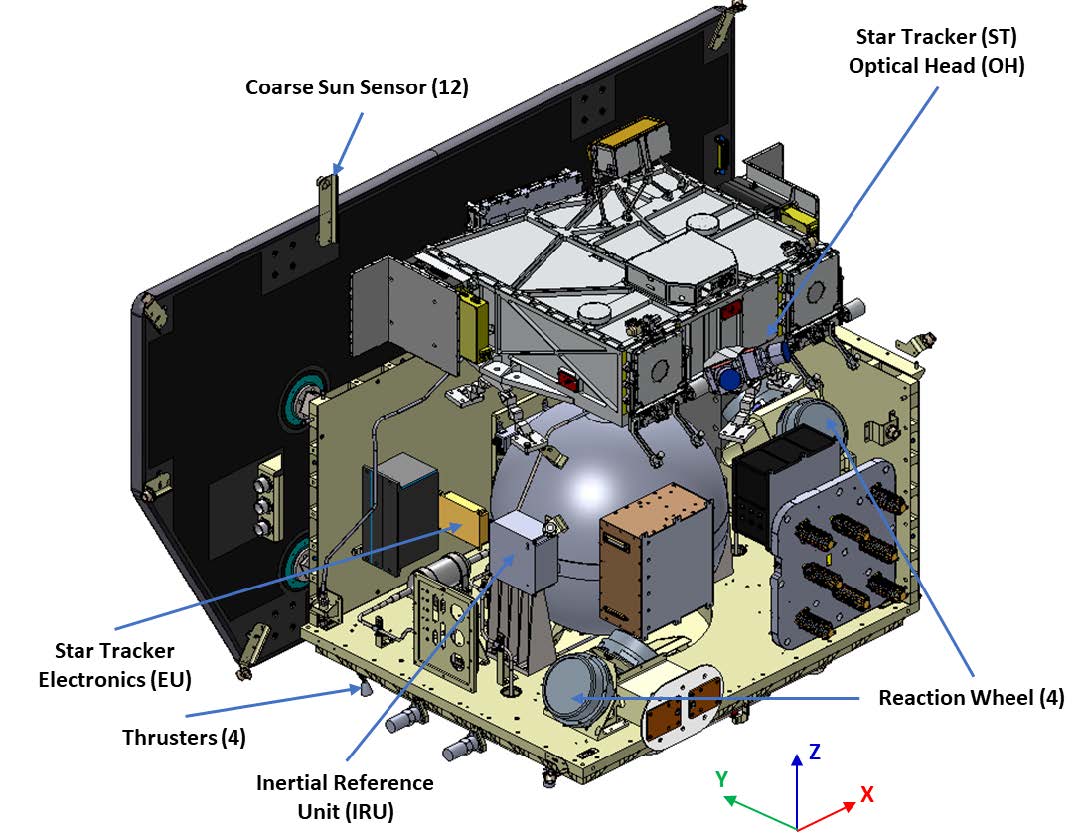}
\caption{The Carruthers spacecraft.}
\label{fig:spacecraft}
\end{figure}

The Carruthers spacecraft (Figure~\ref{fig:spacecraft}) provides a stable, power-positive, and thermally controlled platform for continuous ultraviolet imaging of Earth’s exosphere from a halo orbit about the Earth--Sun L1 point. The spacecraft was developed and integrated by BAE Systems (formerly Ball Aerospace), leveraging heritage from the Small Satellite Evolve product line missions.

The design emphasizes simplicity and operational margin consistent with NASA Class~D mission objectives and the requirement for long-duration autonomous operations. A baseline small-satellite bus architecture was tailored to support the Earth--Sun L1 staring mission geometry and the continuous nadir-pointing science mode required for geocoronal imaging.

Carruthers was launched on 24 September 2025 as a rideshare payload on the IMAP mission mounted to an ESPA-Grande ring. Following separation from the launch vehicle, the spacecraft operates as an independent observatory.

\subsection{Structural Configuration and Payload Accommodation}

The spacecraft employs a compact aluminum primary structure with machined decks and honeycomb panels forming a stiff, low-mass bus. The payload is mounted to a dedicated payload deck using precision interfaces designed to maintain optical alignment and thermal isolation.

The overall configuration accommodates the GeoCoronal Imager field-of-view requirements, star tracker lines of sight, antenna coverage, and solar array illumination without obstruction during nominal observing attitudes.

\subsection{Attitude Determination and Control}

The attitude determination and control system provides precise pointing knowledge and stability to support continuous Earth-centered imaging. Attitude knowledge is derived primarily from two Sodern star tracker optical heads mounted near the instrument boresight, supplemented by coarse sun sensors and an inertial reference unit.

Spacecraft attitude control is provided by a reaction wheel assembly arranged to deliver full three-axis control authority. Thrusters are used for momentum management, orbit correction maneuvers, and initial detumble following separation from the launch vehicle if the tip-off rate exceeds the capability of the reaction wheel array. During nominal science operations, small propulsive maneuvers are performed weekly to reduce  momentum stored in the reaction wheels.

\subsection{
Power, Propulsion, and Thermal Design}

Electrical power is generated by a body-mounted solar array and distributed through a direct-energy-transfer architecture with heritage charge-control algorithms. Energy storage is provided by a lithium-ion battery sized to support science operations, communications, and maneuvering with margin.

Orbit insertion, halo-orbit maintenance, and momentum unloading are accomplished using a blowdown monopropellant propulsion system with four 5-N thrusters. The propulsion system provides sufficient $\Delta V$ margin to support both the baseline two-year mission and potential extended operations.

Thermal control is achieved through a combination of passive insulation, radiative surfaces, and heater-controlled components. Thermal analysis indicates that all spacecraft and payload components remain within allowable temperature limits throughout the mission. 

The observatory concept of operations and archeticture is designed to minimize thermal gradients in the payload to maintain stability of the optics and detectors.   By default is instrument is expected to be operated with the -Z component of the observatory frame of reference intersecting the Sun-Earth line, with the payload effectively shaded by the spacecraft body and solar panel.  Illumination of the +X or -X panels is also an option.  Illumination of the +Z face is avoided, except during flat field observations, to reduce thermal input to the top of the GCI instrument.   Around the halo orbit, the thermal design must accept continuous illumination of the -Z, +X or -X faces of the observatory up to 28$^\circ$, while the power system must accept continuous pointing of the solar panel away from the sun up to 28$^\circ$. The design trade space balanced these competing requirements, providing margin in both power and thermal systems in nadir pointing. This approach enables periods of pointing at higher sun-angles for stellar calibration and views around the perimeter of the exosphere.

\subsection{Telecommunications and Data Handling}

Carruthers communicates with Earth using an S-band telecommunications system supporting encrypted command uplink and telemetry and science data downlink via the NASA Deep Space Network (DSN). Science data are returned at rates up to 1~Mbit~s$^{-1}$ during scheduled DSN contacts.

Command and data handling functions are implemented within a Broad Reach avionics unit hosting a radiation-tolerant processor and solid-state data storage. The avionics architecture supports autonomous fault protection, safe-mode recovery, and flexible mission operations.

\begin{table}
\caption{Summary of key spacecraft and observatory performance metrics.}
\label{tab:spacecraftperformance}
\centering
\begin{tabular}{lllll}
\hline
Parameter & Value & Requirement & Margin & Units \\
\hline
Observatory Dry Mass & 206.2 & 275 & 25\% & kg \\
Observatory Wet Mass & 240.9 & 320 & 24.7\% & kg \\
$\Delta V$ Capability & 221 & 155 & 27\% & m/s \\
Science Mode Power & 194.5 & 268.7 & 28\% & W \\
Pointing Stability & 13.92 & 25 & 79.6\% & arcsec \\
Pointing Knowledge & $\sim$16 & 36--100 & $>$100\% & arcsec \\
Housekeeping Data Rate & 172 & 14 & 1 day margin & MB/day \\
Science Data Rate & 170 & 10 & 11 days margin & MB/day \\
Nominal Downlink Rate & 0.9 & --- & 4.7 dB & Mbps \\
Weekly Downlink Time Required & 5.9 & 7.1 & 41\% & hr/week \\
Available Downlink Time & 12 & --- & --- & hr/week \\
\hline
\end{tabular}
\end{table}

\section{Mission Operations}

Mission operations for the Carruthers Geocoronal Observatory are implemented through a distributed operational architecture consisting of the Mission Operations Center (MOC) at the University of California, Berkeley; the Science Data and Operations Center (SDOC) at the University of Illinois Urbana--Champaign; and a dedicated Flight Dynamics and Navigation (FDNAV) team responsible for trajectory design, orbit determination, and maneuver planning. Together these elements support continuous science operations from a halo orbit about the Earth--Sun L1 point and were fully exercised through Mission Scenario Tests and Operational Readiness Tests prior to launch.

\subsection{Mission Operations Concept and Ground Architecture}

The Carruthers Mission Operations System is designed to support largely autonomous spacecraft operations with human oversight during scheduled ground contacts. Nominal science operations include multiple Deep Space Network (DSN) passes per week providing approximately twelve hours of downlink capacity and margin for command uplink, telemetry return, and science data transmission.

The Mission Operations Center at UC Berkeley is responsible for real-time command and control, spacecraft health and safety monitoring, command sequence generation, and anomaly response. The MOC utilizes the GALAXY command and control system with heritage from previous NASA Heliophysics missions operated at Berkeley.

Mission operations are staffed by UC Berkeley flight controllers and supported by BAE mission controllers and subsystem engineers through observatory checkout and transition to Science Phase E. A team of flight dynamics specialists at Berkeley provides navigation support and designs each of the orbit shaping maneuvers, LOI, and all subsequent stationkeeping maneuvers. All primary and contingency procedures were validated through mission rehearsals conducted prior to launch.

\subsection{Commanding, Monitoring, and Anomaly Response}

Commanding is performed using time-tagged command sequences generated by the MOC and uploaded to onboard command storage memory. These sequences support science observations, calibration activities, communications configuration, and maneuver execution. After observatory checkout, real-time commanding is expected primarily for contingency response or adjustment of payload parameters that require prompt confirmation of any changes in performance.

Spacecraft telemetry is monitored during ground contacts and reviewed post-pass using automated limit checking and trending tools. The Berkeley Emergency and Anomaly Response System provides automated notification to on-call operators in the event of limit violations or unexpected spacecraft behavior. The spacecraft is designed to safe itself if a critical limit is determined to be exceeded, first by transitioning to sun-pointing with a slow roll, and if necessary to safe mode where the payload it powered down. Flight controllers are trained to act as first responders, either placing the observatory into a safe configuration before detailed anomaly investigation, or bringing up key communications systems in order to downlink stored spacecraft data that record the anomaly. 

\subsection{Flight Dynamics and Navigation Operations}

Flight dynamics and navigation activities are conducted by a dedicated team using tools based on NASA’s MONTE trajectory analysis framework. Navigation responsibilities include orbit determination, maneuver planning, station keeping, fuel bookkeeping, and conjunction assessment.

Carruthers operates in a Class~I halo orbit about the Earth--Sun L1 point with a nominal period of approximately 178~days. Station keeping maneuvers are typically performed every 60--90~days. Maneuver strategies target zero $X$-velocity at future $X$--$Z$ plane crossings (GCI coordinates) and draw on heritage approaches flown on the THEMIS and ARTEMIS missions.

Orbit determination solutions are generated using DSN tracking data processed through a MONTE-based filter. Navigation products, including predictive and definitive ephemerides, maneuver execution products, and attitude profiles, are delivered regularly to the MOC and SDOC. Definitive orbit solutions are generated monthly following sufficient tracking data accumulation to reconstruct maneuvers. Navigation solutions are also provided as ancillary data products to the science data processing pipeline and delivered to NASA’s MADCAP system to support conjunction assessment and collision avoidance planning.

\subsection{Science Operations and MOC--SDOC Interface}

The Science Data Operations Center at the University of Illinois Urbana--Champaign is responsible for science planning, data processing, product validation, and data archival.

\begin{figure}
\centering
\includegraphics[width=\linewidth]{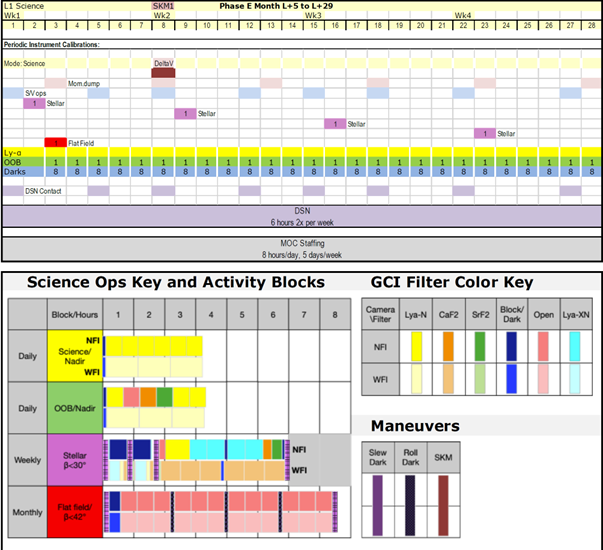}
\caption{A nominal plan for science activity during the baseline 2-year science mission.}
\label{fig:activity}
\end{figure}

Science observation plans are generated by the SDOC as sequences of activity blocks defining pointing geometry, filter selection, integration times, and observation cadence. These plans are delivered to the MOC where they are converted into spacecraft command sequences and validated against spacecraft operational constraints.  A nominal science plan for the two-year baseline mission, beginning on 1 March 2026, is shown in (Figure~\ref{fig:activity}).  

Raw Level~0 telemetry and associated ancillary data are transferred from the MOC to the SDOC following each ground contact. The SDOC processes these data through a multi-stage pipeline producing calibrated images (Level 1) and derived geophysical products including global hydrogen density distributions (Level 2).

\subsection{Operations Readiness and Verification}

The Carruthers mission operations concept was validated through a campaign of Mission Scenario Tests and Operational Readiness Tests that exercised commissioning procedures, routine science operations, and off-nominal scenarios. These tests verified the readiness of the mission operations system, ground systems, and operations team to support launch, commissioning, and sustained science operations at the Earth--Sun L1 point.

\section{Data Products and Archiving}

The Carruthers mission produces a hierarchy of calibrated imaging and derived geophysical data products designed to support both focused scientific investigations and long-term community use. Data processing, validation, and distribution are performed by the Science Data Operations Center at the University of Illinois Urbana--Champaign in coordination with the Mission Operations Center and flight dynamics team.

Raw telemetry packets are processed into Level~0 products following each ground contact and subsequently converted into calibrated Level~1 images through radiometric correction, flat-fielding, background subtraction, and geometric registration. Higher-level products include line-of-sight integrated hydrogen brightness and three-dimensional hydrogen density distributions retrieved using forward modeling and inversion techniques.

All data products are accompanied by complete metadata including spacecraft ephemerides, pointing history, calibration versions, and processing provenance. Navigation solutions and definitive ephemerides produced by the flight dynamics team are incorporated into the science data stream to ensure accurate spatial and temporal registration.

Validated data products are archived and made publicly available through NASA’s Space Physics Data Facility in standard heliophysics data formats consistent with NASA open-data policies. Documentation and software tools supporting data interpretation are released concurrently to facilitate community use.

\section{Summary}

The NASA Carruthers Mission is one of two missions ever flown to sun-Earth L1 orbit specifically designed to obtain continuous images of Earth \citep{zastrow15}. The locale is ideal for terrestrial observations with a straightforward concept of operations, but requires capability that differs significantly from the more often-flown space weather monitoring systems. The Carruthers observatory possesses capability with significant margin to requirements to provide flexibility in routine science operations and calibration maneuvers. Designed for a 2-year mission, the consumables available provide for a potential extended mission of more than a decade.

\section*{Acknowledgments}

The authors acknowledge the contributions of the Carruthers science and engineering teams and support from the NASA Heliophysics Division and the Solar Terrestrial Probes Program. The Carruthers award number is NASA 80GSFC21C0038. The authors acknowledge Chat-GPT for developing a framework and first draft of this paper.

\bibliographystyle{spbasic}
\bibliography{carruthers_refs}

\end{document}